\documentclass[sigconf]{acmart}

\setcopyright{acmlicensed}
\copyrightyear{2018}
\acmYear{2018}
\acmDOI{XXXXXXX.XXXXXXX}

\acmJournal{JACM}
\acmVolume{37}
\acmNumber{4}
\acmArticle{111}
\acmMonth{8}

\usepackage{quiver}
\usepackage{macros}
\usepackage{savoirs}

\begin{document}

\title{A cartesian closed fibration of higher-order regular languages\vspace{.2em}
}

\author{Paul-Andr{\'e} Melli{\`e}s}
\email{mellies@irif.fr}
\orcid{0000-0001-6180-2275}
\affiliation{%
    \institution{CNRS, Universit\'e Paris Cit\'e, INRIA}
    \country{France}
}

\author{Vincent Moreau}
\email{vincent.moreau@taltech.ee}
\orcid{0009-0005-0638-1363}
\affiliation{
    \institution{Tallinn University of Technology}
    \country{Estonia}
}

\renewcommand{\shortauthors}{Melliès et Moreau}

\begin{abstract}
    We explain how to construct in two different ways
    a cartesian closed fibration of higher-order regular languages
    in the sense of Salvati.
    %
    In the first construction, we use fibrational techniques
    to derive the cartesian closed fibration from
    the various categories of regular languages
    of $\lambda$-terms associated to finite sets of ground states.
    In the second construction, we take advantage of the recent notion
    of profinite $\lambda$-calculus to define the cartesian closed fibration by a change-of-base from the fibration
    of clopen subsets over the category of Stone spaces, using an elegant
    idea coming from Hermida.
    We illustrate the expressive power of the cartesian closed fibration
    by generalizing the notion of Brzozowski derivative
    to higher-order regular languages,
    using an Isbell-like adjunction in the sense of Melli{\`e}s and Zeilberger.
    %
    %
\end{abstract}



\received{20 February 2007}
\received[revised]{12 March 2009}
\received[accepted]{5 June 2009}


\maketitle

\section{Introduction}
\label{sec:introduction}
The class of regular languages over finite words
is closed under two basic constructions:
inverse images along homomorphisms and
Brzozowski derivatives.
On the one hand, given two finite alphabets~$\Sigma$ and~$\Gamma$
and a homomorphism $h:\Sigma^{*}\to\Gamma^{*}$,
the inverse image
$$h^{-1}(L) = \{ \, w\in\Sigma^{*} \,\, | \,\, h(w)\in L\, \}$$
of a regular language~$L$ is also regular.
On the other hand, Brzozowski derivatives
\begin{align*}
    a \backslash L \quad & := \quad \{w \in \Sigma^* \mid aw \in L\}
    \\
    L \slash a \quad     & := \quad \{w \in \Sigma^* \mid wa \in L\}
\end{align*}
In fact, the Brzozowski construction is much more general,
since the languages
\begin{align*}
    L' \backslash L \quad & := \quad \{w \in \Sigma^* \mid \forall w'\in \Sigma^{*}, \, w'\in L'\Rightarrow w'w \in L\}
    \\
    L \slash L' \quad     & := \quad \{w \in \Sigma^* \mid \forall w'\in \Sigma^{*}, \, w'\in L'\Rightarrow ww' \in L\}
\end{align*}
are regular when $L$ and $L'$ are regular languages.
Note that the derivatives $a \backslash L$ and $L \slash a$ are recovered
by instantiating $L' \backslash L$ and $L' \slash L$ at the singleton language~$L'=\{a\}$
which is regular.

At the same time, the class of regular languages over finite words
is closed under concatenation, Kleene star, and direct image along homomorphisms.
This means that given two finite alphabets~$\Sigma$ and~$\Gamma$
and a homomorphism~$h:{\Sigma}^{\ast}\to{\Gamma}^{*}$,
the direct image
$$h(L) = \{ \, h(w)\in\Gamma^{*} \,\, | \,\, w\in L\, \}$$
of a regular language~$L$ is also regular.
The fact that regular languages are closed under direct image along homomorphisms
is a remarkable property, which has been recognized in the literature as the fundamental
reason for the correspondence with MSO formulas \cite{Bojaczyk2023}.

\medbreak

\noindent\textbf{Regular languages of tuples of words.}
An interesting observation is that the direct and inverse image preservation properties
cannot hold together in an extended and speculative language theory,
where one would not just consider regular languages
$$
    L\subseteq \Sigma^{\ast}
$$
of finite words, but also regular languages
$$L\subseteq \Sigma^{\ast}\times\Sigma^{\ast}$$
of pairs of words.
Indeed, the diagonal function
$$
    \Delta : \Nat \to \Nat\times\Nat
$$
would transport in that case
the total regular language $L=\Nat$ over $\Nat\cong\{a\}^{*}$
to the language
$$
    \Delta (\Nat) \quad = \quad \{(n,n)\mid n\in\Nat\}.
$$
In turn, the inverse image of $\Delta (\Nat)$
along the function
$$
    w\mapsto (|w|_a,|w|_b) \quad : \quad \{a,b\}^{*} \to \Nat\times\Nat
$$
which associates to every word $w$ the number $|w|_a$ and $|w|_b$
of letters $a$ and $b$, is the language
$$
    L \,\, = \,\, \{ \, w\in\{a,b\}^{*} \, \mid \, |w|_a=|w|_b \,\}
$$
which is not regular.
For that reason, the language $\Delta (\Nat)$
describing the diagonal in $\Nat\times\Nat$ would not
be regular, hence the direct image preservation property
would not hold in the extended and speculative
language theory for tuples of finite words we have in mind.

\medbreak

\noindent\textbf{Regular languages of simply-typed $\lambda$-terms.}
Fifteen years ago, the traditional notion of regular language
of finite words was conservatively extended by Salvati~\cite{DBLP:conf/wollic/Salvati09}
to a notion of higher-order regular language of
simply-typed $\lambda$-terms.
Suppose given a simple type~$A$ generated by the grammar
$$
    A,B \quad ::= \quad A\times B \mid A\Rightarrow B \mid 1 \mid \tyo
$$
with unit type~$1$ and a unique base type~$\tyo$.
A \emph{higher-order language}~$L$
of type~$A$ is defined as a subset $L\subseteq\Tm(A)$
of closed $\lambda$-terms
of type~$A$, up to $\beta\eta$-equivalence.
%
%
The notion of {regular} language of type~$A$
relies on the standard interpretation
\begin{center}
    $\sem{A}{Q} \hspace{.5em} \in \hspace{.5em} \FinSet$
\end{center}
of the simple type~$A$ in the cartesian closed
category~$\FinSet$ of finite sets, where
the base type~$\tyo$ is interpreted as a finite set~$Q$.
%
For a given finite set~$Q$,
a higher-order language~$L\subseteq\Tm(A)$
is called $Q$-\emph{regular} when there exists
a finite subset
\begin{center}
    $F \hspace{.5em} \subseteq \hspace{.5em} \sem{A}{Q}$
\end{center}
such that
$$L \hspace{.5em} = \hspace{.5em} \big\{ \, M \in \Tm(A) \, \mid \, \sem{M}{Q} \in F \,\big\}.$$
Here, in analogy with traditional automata theory,
one should think of the finite set~$Q$ as a set
of ground states, and of~$F$
as a subset of accepting higher-order states of type~$A$.

Note that for every finite set~$Q$,
the standard interpretation
of~$A$ as the finite set~$\sem{A}{Q}$
induces a partition of $\Tm(A)$
into a finite number of disjoint components,
where each component is a nonempty
subset of closed $\lambda$-terms $M\in\Tm(A)$
with the same interpretation:
$$
    \sem{M}{Q} \in \sem{A}{Q}.
$$
Hence, a higher-order language $L$ of type~$A$
is $Q$-regular precisely when it is the finite union
of a number of these disjoint components.
A  higher-order language~$L\subseteq\Tm(A)$
of type~$A$ is then called \emph{regular}
when there exists a finite set~$Q$
such that $L$ is $Q$-regular, see~\cite{DBLP:conf/wollic/Salvati09}.

\medbreak

\noindent
\textbf{The Church encoding of finite words.}
As mentioned above,
the definition of regular higher-order languages
formulated by Salvati
provides a \emph{conservative} extension of the usual notion
of regular language of finite words.
Every finite alphabet
$$\Sigma=\{a_1,\dots,a_n\}$$
can be seen as a context
of the simply typed $\lambda$-calculus:
$$
    a_1:\tyo\Rightarrow\tyo
    \, , \,
    \dots
    \, , \,
    a_n: \tyo\Rightarrow\tyo
$$
where every letter~$a_i\in\Sigma$ is declared
as a function of type $\tyo\Rightarrow\tyo$
from the base type~$\tyo$ to itself.
The idea of the Church encoding is to translate every finite word
$$w \,\, = \,\, a_{\sigma 1}\dots a_{\sigma k}$$
where $\sigma:\{1,\dots,k\}\to\{1,\dots,n\}$ is a function
from positions to letter indices,
as the $\lambda$-term
$$[w]  \,\, = \,\, a_{\sigma k} \circ \dots \circ a_{\sigma 1}$$
obtained by composing
the variables $a_{\sigma i}$ declared as functions
in the context.
One obtains in that way a $\lambda$-term
of type~$\tyo\Rightarrow\tyo$ in that context
or typed alphabet:
$$
    a_1:\tyo\Rightarrow\tyo
    \, , \,
    \dots
    \, , \,
    a_n: \tyo\Rightarrow\tyo
    \,\, \vdash \,\, [w] \,\, : \,\, \tyo\Rightarrow\tyo$$
The composition of a sequence of~$k$ $\lambda$-terms
$f_1,\dots,f_k$ is itself defined in the $\lambda$-calculus
as the following $\lambda$-term:
$$
    f_{1} \circ \dots \circ f_{k} \quad ::= \quad
    \lambda x. f_{1} ( f_{2} ( \dots f_{k} (x)\dots ))
$$
Note in particular that the empty word
is encoded as the identity function,
noted $\lambda x.x$ in the $\lambda$-calculus:
$$
    a_1:\tyo\Rightarrow\tyo
    \, , \,
    \dots
    \, , \,
    a_n: \tyo\Rightarrow\tyo
    \,\, \vdash \,\,\lambda x. x \,\, : \,\, \tyo\Rightarrow\tyo$$
In that way, every finite word~$w\in\Sigma^{*}$
can be seen as the $\lambda$-term
$$
    \vdash \,\, \lambda a_1.\dots.\lambda a_n. [w] \,\, : \,\, \Words{\Sigma}$$
obtained by currification of the $\lambda$-term~$[w]$,
where~$\Words{\Sigma}$ is defined as the simple type
$$
    \Words{\Sigma}
    \,\,:=\,\,
    \underbrace{(\tyo\Rightarrow\tyo)}_{\mbox{type of $a_1$}}
    \Rightarrow\dots\Rightarrow
    \underbrace{(\tyo\Rightarrow\tyo)}_{\mbox{type of $a_n$}}\Rightarrow
    \underbrace{(\tyo\Rightarrow\tyo)}_{\mbox{type of $[w]$}}
$$
The Church encoding of finite words
into simply typed $\lambda$-terms is one-to-one
in the sense that a finite word $w\in\Sigma^{*}$ is the same thing
as a closed $\lambda$-term of type~$\Words{\Sigma}$,
up to $\beta\eta$-conversion:
%
$$
    \Sigma^{*}
    \hspace{.5em} \cong  \hspace{.5em}
    \Tm(\Words{\Sigma}).
$$
By conservative extension, one means that
the regular higher-order languages
of type $\Words{\Sigma}$
coincide with the regular languages of finite words
on the alphabet $\Sigma$
in the traditional sense,
after Church encoding of finite words
on the alphabet~$\Sigma$ into simply typed $\lambda$-terms
of type~$\Words{\Sigma}$.
As a matter of fact, this basic but important observation motivates the definition
of regular higher-order language in the original paper by Salvati~\cite{DBLP:conf/wollic/Salvati09}.

\medbreak

\noindent
\textbf{Product and arrow types, inverse images and Brzozowski derivatives on higher-order languages.}
Our main ambition in the present paper
is to establish that the class of regular higher-order languages
is closed under two basic constructions: product types and
arrow types (Theorem A), as well as inverse image along a $\lambda$-term (Theorem B).

%
%
We describe the constructions.
Suppose given two regular languages
$$
    L_A\subseteq\Tm(A)
    \quad\quad
    L_B\subseteq\Tm(B)
$$
The product and arrow type constructions
$$
    L_A\times L_B\,\,\subseteq\,\,\Tm(A\times B)
    \quad\quad
    L_A\Rightarrow L_B \,\, \subseteq \,\,\Tm(A\Rightarrow B)
$$
are defined as follows:
\begin{center}
    $L_A\times L_B \,\, = \,\, \big\{\,\, (M,N)\,\in\Tm(A\times B)\, \mid \,\, M\in L_A, N\in L_B \,\, \big\}$
\end{center}
\begin{center}
    $L_A\Rightarrow L_B \,\, = \,\, \big\{\, M\in\Tm(A\Rightarrow B) \, \mid \, \forall P, \,
        P\in L_A\Rightarrow MP\in L_B \, \big\}$
\end{center}

\noindent
We establish in the paper that

\medbreak
\begin{center}
    \fcolorbox{littleblue}{paleyellow}{
        \hspace{-1.2em}
        \begin{tabular}{l}
            \vspace{-1.8em}
            \\
            \textbf{Theorem A.}
            \\
            The higher-order languages~$L_A\times L_B$
            and $L_A\Rightarrow L_B$
            \\
            are regular when $L_A$ and $L_B$ are regular.
            \\
            \vspace{-1.8em}
        \end{tabular}
        \hspace{-1.2em}}
\end{center}
\medbreak



\noindent
Suppose given a closed $\lambda$-term
$$\vdash \hspace{.5em} M \, : \, A\Rightarrow B$$
and a higher-order language~$L$ of type~$B$.
The inverse image
\begin{center}
    $M^{-1} (L) \,\, \subseteq \,\, \Tm(A)$
\end{center}
is the higher-order language of type~$A$ defined as
\begin{center}
    $M^{-1} (L) \,\, = \,\, \big\{\, P\in\Tm(A)  \, \mid \, MP\in L \, \big\}$
\end{center}

\noindent
We establish in the paper that

\medbreak
\begin{center}
    \fcolorbox{littleblue}{paleyellow}{
        \hspace{-1.2em}
        \begin{tabular}{l}
            \vspace{-1.8em}
            \\
            \textbf{Theorem B.}
            \\
            The higher-order language~$M^{-1}(L)$ is regular
            \\
            when $L$ is regular.
            \\
            \vspace{-1.8em}
        \end{tabular}
        \hspace{-1.2em}}
\end{center}
\medbreak


\noindent
Combined together, the theorems A. and B. have a number
of important consequences.
First of all, the intersection~$L\cap L'$ of two languages~$L$ and~$L'$
of type~$A$ is equal to the inverse image
$$
    L\cap L' = \Delta_A^{-1}(L\times L')
$$
of the regular language $L\times L'$ along the $\lambda$-term $\Delta_A$
$$
    \begin{tikzcd}
        \vdash \quad {\Delta_A} \,\, : \,\, A\Rightarrow (A\times A)
    \end{tikzcd}
$$
From this follows that $L\cap L'$ is regular
when the languages~$L$ and~$L'$ are regular of type~$A$.
The set~$\Reg{(A)}$ of regular languages of type~$A$
is moreover closed under complementation
since the set~$\Reg_{\,Q}{(A)}$ of $Q$-regular languages
of type~$A$ is closed under complementation for obvious reasons.
The set~$\Reg{(A)}$ also contains the empty set as well as total set $\Tm(A)$.
This provides a conceptual proof of the subtle and well-known fact
that $\Reg{(A)}$ defines a boolean algebra.
%

\medbreak

\noindent
The two theorems A. and B. also provide enough structure
to extend Brzozowski derivatives to higher-order languages,
in the following way.
Suppose given a closed $\lambda$-term~$M$ of type
$$
    \vdash \quad M \,\, : \,\, A \times B \, \Rightarrow \, C
$$
and a regular language~$L_C$ of type~$C$.
%
We establish in \S\ref{sec:logic-of-regular-languages}
that the higher-order language of type~$B$
$$
    L_C \slash L_A \, = \,
    \big\{\,N_B : B \mid \forall N_A \in L_A, \text{ we have } M\ (N_A, N_B) \in L_C\,\big\}
$$
is regular when the language~$L_A$ of type~$A$ is regular~;~and symmetrically, that the higher-order language of type~$A$
$$
    L_B \backslash L_C \, = \,
    \big\{\, N_A : A \mid \forall N_B \in L_B, \text{ we have } M\ (N_A, N_B) \in L_C\,\big\}
$$
is regular when the language~$L_B$ of type~$B$ is regular.
%
%
We recover the usual Brzozowski derivative on languages of finite words
by instantiating the construction at the types $A,B,C=\Church{\Sigma}$
and at the $\lambda$-term~$M=concat$ defined as
%
$$\vdash \quad concat \,\, : \,\, \Words{\Sigma}\times \Words{\Sigma}\to \Words{\Sigma}$$
which transports the Church encoding of a pair $(u,v)\in\Sigma^{\ast}\times\Sigma^{\ast}$ of two words
to the Church encoding of their concatenation~$u\cdot v\in\Sigma^{\ast}$, see \S\ref{sec:logic-of-regular-languages} for details.
%

\medbreak

\noindent\textbf{A cartesian closed fibration of regular languages.}
The two theorems A. and B. describe fundamental closure properties
of regular languages of $\lambda$-terms,
%
which were inspired
by the idea that higher-order languages~$L$ of type~$A$
should be understood
as \emph{refinement types} $L\sqsubset A$ of the underlying simple type~$A$,
and therefore studied in the fibrational and type-theoretic formalism
developed in~\cite{MelliesZeilberger15}.

To that purpose, we consider the cartesian closed category~$\Lam$
freely generated by one object~$\tyo$, whose objects are simple types~$A$
on the base type~$\tyo$, and whose morphisms are simply typed $\lambda$-terms,
considered up to $\beta\eta$-conversion.
We then define the category~$\Reg$ whose objects are pairs $(A,L)$
of a simple type~$A$ and of a regular language~$L$ of type~$A$.
A morphism
\begin{center}
    \begin{tikzcd}[column sep = 1.2em]
        M \quad : \quad (A,L_A)\arrow[rr] && (B,L_B)
    \end{tikzcd}
\end{center}
is a $\lambda$-term~$M$ of type~$A\Rightarrow B$
such that
$$
    \forall P\in\Tm(A), \quad P\in L_A \, \Rightarrow \, MP\in L_B.
$$
Note that this property can be equivalently stated as $L_A\subseteq M^{-1}(L_B)$.
The statement of \textbf{Theorem A.} can be reformulated (and strengthened)
by observing that the constructions $L_A\times L_B$ and
$L_A\Rightarrow L_B$ equip the category~$\Reg$
with the structure of a cartesian-closed category,
strictly preserved by the functor
$$
    \begin{tikzcd}
        p \quad : \quad \Reg \arrow[rr] && \Lam
    \end{tikzcd}
$$
which transports every object~$(A,L)$ to its underlying type~$A$.
The statement of \textbf{Theorem B.} can be reformulated (and strengthened)
by observing that the functor~$p$ is a Grothendieck fibration.

\medbreak

From this follows that \textbf{Theorems A. and B.} are implied
and at the same time refined by \textbf{Theorem C.} which is
the main focus and contribution of the paper:


%

%


\medbreak
\begin{center}
    \fcolorbox{littleblue}{paleyellow}{
        \hspace{-1.2em}
        \begin{tabular}{l}
            \vspace{-1.8em}
            \\
            \textbf{Theorem C.} \, [Thms.~\ref{thm:regcolim-ccc-fibration} and~\ref{thm:reg-stone-ccc-fibration} in this paper]
            \\
            The functor $p:\Reg\to\Lam$ is
            a cartesian-closed fibration.
            \\
            \vspace{-1.8em}
        \end{tabular}
        \hspace{-1.2em}}
\end{center}
\medbreak

\medbreak

\noindent
Theorem C. is established in two different ways in the paper.

\medbreak

\noindent\textbf{First proof: a fibrational approach.}
Our first proof of the theorem
is described in \S\ref{sec:a-bifibration-of-Q-regular} and \S\ref{sec:fibered-adjunctions}.
%
We start by constructing for every finite set~$Q\in\FinSet$,
a cartesian-closed bifibration
$$
    \begin{tikzcd}
        p_Q \quad : \quad \Reg_{Q} \arrow[rr] && \Lam
    \end{tikzcd}
$$
from the category $\Reg_{Q}$ of $Q$-regular languages
to the category~$\Lam$, see Prop.~\ref{prop:RecQtoLam}.
We then use the notion of reflective morphism of a bifibration (Def.~\ref{def:reflective-morphism})
and the Frobenius reciprocity principle (see Thm.~\ref{thm:reflective-change-of-base})
to establish our main theorem (Thm.~\ref{thm:regcolim-ccc-fibration}).
%

\medbreak

\noindent\textbf{Second proof: a profinite approach.}
Our second proof of the theorem appears in \S\ref{sec:fibration-topological-predicates}
and \S\ref{sec:reg-through-profinite-lambda-terms}.
Its conceptual simplicity illustrates the benefits
of using the notion of \emph{profinite $\lambda$-calculus}
recently introduced in \cite{entics:12280}.
The proof starts with the construction of a bifibration
$$
    \begin{tikzcd}
        \Clopen \arrow[rr] && \Stone
    \end{tikzcd}
$$
from the category $\Clopen$ whose objects are pairs $(X,U)$
consisting of a Stone space~$X$ and a clopen subset $U\subseteq X$.
The profinite $\lambda$-calculus defines a functor
\[
    \ProTm
    \quad:\quad
    \Lam
    \ \longto\
    \Stone
\]
which associates to every simple~$A$ its Stone space
of profinite $\lambda$-terms~$\ProTm(A)$,
which can be seen as a compactification of~$\Tm(A)$.
The cartesian-closed fibration of regular languages $p:\Reg\to\Lam$
is then recovered by a change-of-base, defined by the categorical pullback:
\[
    \begin{tikzcd}[ampersand replacement=\&]
        \Reg \& \Clopen \\
        \Lam \& \Stone
        \arrow[from=1-1, to=1-2]
        \arrow[from=1-1, to=2-1]
        \arrow["\lrcorner"{anchor=center, pos=0.125}, draw=none, from=1-1, to=2-2]
        \arrow[from=1-2, to=2-2]
        \arrow["\ProTm", from=2-1, to=2-2]
    \end{tikzcd}
\]
The proof that $p:\Reg\to\Lam$ defines a cartesian-closed fibration (Thm.~\ref{thm:reg-stone-ccc-fibration})
relies on an argument adapted from Hermida's PhD thesis.

\medbreak

\noindent\textbf{Regular languages of tuples of finite words, and the notion of open maps.}
Among its numerous benefits, the notion of regular higher-order language
provides us with a conservative extension of regular languages of \emph{finite words}
$$w\,\, \in \,\, \Sigma^{\ast}$$
to regular languages of \emph{tuples of finite words}
$$(w_1,\dots,w_n) \,\, \in \,\, \Sigma_1^{\ast}\times\dots\times\Sigma_n^{\ast}$$
Typically, a regular language
$$L\subseteq \Sigma^{\ast}\times\Sigma^{\ast}$$
of pairs of finite words on the finite alphabet~$\Sigma$ is defined
as a regular higher-order language
$$L \subseteq \Tm(\Words{\Sigma}\times\Words{\Sigma})$$
of type $\Words{\Sigma}\times\Words{\Sigma}$.
%
%
%
As we discussed earlier in the introduction,
this establishes that regular higher-order languages
are \emph{not preserved} by direct image for deep structural reasons,
related to the essence of regularity and recognizability by finite means,
such as finite automata.
%
Indeed, consider the Church encoding
$$
    \Nat = (\tyo\Rightarrow\tyo) \Rightarrow (\tyo\Rightarrow\tyo)
$$
of the set of natural numbers.
There exists a $\lambda$-term
$$
    \begin{tikzcd}[column sep = 1em]
        \textit{counter} \quad : \quad \Words{\{a,b\}} \arrow[rr] && \Nat\times\Nat
    \end{tikzcd}
$$
which associates to every finite word~$w$ on the alphabet~$\{a,b\}$
the pair $(|w|_{a},|w|_{b})$ of numbers of letters~$a$ and~$b$ appearing in the word~$w$.
Here, $\Nat$ is defined as the Church encoding
of the set of natural numbers:
$$
    \Nat = (\tyo\Rightarrow\tyo) \Rightarrow (\tyo\Rightarrow\tyo)
$$
From this follows that:

\medbreak

\noindent
\textbf{Fact.} The higher-order language
$$L \, = \, \{\, (n,n)\mid n\in\Nat \, \} \,\, \subseteq \,\, \Tm(\Nat\times\Nat)$$
is not regular.

\medbreak

\noindent
This shows that the direct image~$L$ of the total language $\Tm(\Nat)$
along the diagonal $\lambda$-term
$$
    \Delta : \Nat \to \Nat\times\Nat
$$
is not a regular higher-order language.

\medbreak

The fact that regular higher-order languages are not necessarily preserved
by direct image was already noticed by Salvati
with a sophisticated counterexample to the possibility of designing a MSO logic
for higher-order regular languages, see \cite{DBLP:conf/wollic/Salvati09}.

\medbreak

An elementary but insightful observation is that a simply-typed $\lambda$-term
$
    \vdash \hspace{.2em} M \,\, : \,\, A \Rightarrow B
$
is \emph{open} in the topological sense precisely when $M$ has the property
that the direct image $M(L)$ of any regular language~$L$ of type $A$
is a regular language of type~$B$, see \S\ref{sec:classification-open}
for a discussion and a number of preliminary observations on open maps.

\medbreak

\noindent
\textbf{Summary of the paper.}
After recalling the notions of cartesian-closed fibrations
and bifibrations in \S\ref{sec:change-of-base},
we give in \S\ref{sec:a-bifibration-of-Q-regular}
and \S\ref{sec:fibered-adjunctions}
our first construction of the cartesian closed fibration
of regular languages.
%
%
We then discuss in \S\ref{sec:fibration-topological-predicates}
the fibration of clopen predicates over the category of Stone space.
We give in \S\ref{sec:reg-through-profinite-lambda-terms}
our second construction of the cartesian closed fibration
of regular languages.
%
We then discuss in \S\ref{sec:logic-of-regular-languages}
an extension of the Brzozowski derivatives to higher-order languages.
%
We then provide a number of preliminary observations
on open maps in \S\ref{sec:classification-open},
and then conclude in~\S\ref{sec:conclusion}.

\medbreak

\noindent
\textbf{Related works.}
Salvati extended the notion of recognizability to the setting of language of $\lambda$-terms in \cite{DBLP:conf/wollic/Salvati09}.
Moreau and \Nguyen recently showed that all the non-posetal, locally finite, well-pointed cartesian closed categories
recognize the same languages of $\lambda$-terms.
The class of regular languages coincide with the higher-order language obtained by generalizing
the approach of Hillebrand and Kanellakis to the higher order \cite{DBLP:conf/lics/HillebrandK96},
and relates it to the \textit{implicit automata} research programme started in \cite{nguyen-pradic-1}, see \Nguyen's PhD thesis \cite{titoPhD}.
This demonstrates the robustness of the class of higher-order regular languages.
and allows us to use recognizability in the category~$\FinSet$ without any loss of generality.

%
A fibrational analysis of the structure underlying Rabin’s Tree Theorem,
stating that languages defined by non-deterministic automata on infinite trees are closed under complement,
has been carried out by Riba
based on tools coming from game semantics and linear logic
\cite{Riba_2020, DBLP:journals/mscs/Riba20}.
This research, at the intersection of logic and denotational semantics,
is closely related to the series of papers by Pradic and Riba
building a Curry-Howard viewpoint for MSO centered on Church's Synthesis,
relying on intuitionistic logic, linear logic, and the Dialectica translation \cite{lmcs:4414, Pradic2018, Pradic2019}.

Several techniques used in this article have been pioneered by Hermida \cite{Hermida1993} and Jacobs \cite{JacobsCLTT} in their approach to logical relations.
Although we focus here on cartesian closed categories,
the generality of Hasegawa's work on gluing for linear logic,
which applies to monoidal closed categories,
was a major source of inspiration~\cite{hasegawa:glueing-cll, DBLP:conf/tlca/Hasegawa99},
as well as uses of such techniques in programming language theory such as Katsumata's work on effects \cite{Katsumata2013}.

The idea that any functor represents a refinement type system comes from the work of Melliès and Zeilberger,
who demonstrate how the fibrational and monoidal closed structures interact,
in particular by recovering multiple constructions coming from separation logic \cite{MelliesZeilberger15}.
These ideas were then later developed in relation to Isbell duality,
a construction from category theory akin to a form of double dualization \cite{DILIBERTI2020106379},
establishing links with polarities in proof theory,
the theory of continuations, and the Day tensor product
in \cite{MelliesZeilberger18}.

Initiated by the work of Pippenger \cite{Pippenger1997} and Almeida \cite{doi:10.1142/2481},
the topological approach to automata theory centered around profinite words now contains a mature corpus of concepts and techniques developed for more than 25 years,
see Pin's survey \cite{pin2009},
which also finds application in other topics such as symbolic dynamics \cite{Almeida2020}.
One \textit{tour de force} of Gehrke, Grigorieff and Pin was to demonstrate that the monoid structure on profinite words given by their concatenation is dual to the residuation operators on the Boolean algebra of regular languages, which is computed by Brzozowski derivative \cite{DBLP:conf/icalp/GehrkeGP08, DBLP:conf/icalp/GehrkeGP10}.
Inspired by this work and the one of Salvati on languages of $\lambda$-terms,
Gool, Melliès and Moreau introduced the profinite lambda-calculus which generalizes in a compositional way the notion of profinite word to the higher order \cite{entics:12280}.
Further studied in Moreau's PhD thesis \cite{moreau:tel-05428993},
the profinite $\lambda$-calculus is the crucial tool of \S\ref{sec:reg-through-profinite-lambda-terms} in this article.

\section{Preliminaries on cartesian closed fibrations and bifibrations}
\ZAP\label{sec:change-of-base}

The notion of fibration has been introduced in \cite{Grothendieck1971}.
We refer to \cite{JacobsCLTT} for the definitions of "cartesian" and "cocartesian" lifts.

\begin{definition}
    \label[definition]{def:fibrations}
    Let $p : \E \to \B$ be a functor.

    A morphism $f : X \to Y$ of $\B$ has ""pullbacks"" if
    for every object $V$ of $\E$ such that $p(V) = Y$,
    there exists a cartesian lifting $\pull{f}(V) \to V$ of $f$ to $\E$.
    We say that $p$ is a ""fibration"" if all morphisms of $\B$ have ""pullbacks"".

    A morphism $f : X \to Y$ of $\B$ has ""pushforwards"" if
    for every object $U$ of $\E$ such that $p(U) = X$,
    there exists a cocartesian lifting $U \to \pull{f}(U)$ of $f$ to $\E$.
    We say that $p$ is an ""opfibration"" if all morphisms of $\B$ have ""pushforwards"".

    A ""bifibration"" is a functor which is both a "fibration" and an "opfibration".

    A functor is ""cartesian closed"" when its domain and codomain categories are cartesian closed,
    and it preserves the cartesian products hom strictly,
    and it is ""cartesian closed"" if the domain and codomain categories are moreover closed,
    and it preserves the internal hom strictly.
\end{definition}

\AP
For a functor $p : \E \to \B$, we follow the philosophy of \cite{MelliesZeilberger15}
and we write $\intro*U \refines X$ to mean that $p(U) = X$,
and we write $\E(X)$ for the fiber of $X$.
We write
\[
    \judge{U}{f}{V}
\]
for the set of morphisms $u : U \to V$ such that $p(u) = f$,
and make extensive use of natural bijections which we write with as reversible rules, i.e. with two horizontal bars.
For example, "pushforwards" and "pullbacks" amount to reversible rules
\[
    \begin{prooftree}
        \hypo{\judge{\push{f}(U)}{g}{V}}
        \infer[double]1{\judge{U}{f;g}{V}}
    \end{prooftree}
    \qquad\text{and}\qquad
    \begin{prooftree}
        \hypo{\judge{U}{f}{\pull{g}(V)}}
        \infer[double]1{\judge{U}{f;g}{V}}
    \end{prooftree}
\]

\begin{definition}
    \ZAP\label[definition]{def:subset}
    We write $\intro*\SubSet$ for the category
    \begin{itemize}
        \item whose objects are pairs~$(X, S)$ of a set~$X$ and a subset~$S \subseteq X$,
        \item whose morphisms~$(X, S) \to (Y, T)$ are set-theoretic functions~$f : X \to Y$
              such that for all $x \in X$,
              \[
                  \text{if}\ x \in S
                  \ ,
                  \quad
                  \text{then}\ f(x) \in T
                  \ .
              \]
    \end{itemize}
    We write $p$ for the functor $\SubSet \to \intro*\Set$ which sends any pair~$(X, S)$ on the set $X$.
\end{definition}

This functor $\SubSet \to \Set$ is a "cartesian closed" "bifibration",
whose "pushforwards" are given by the direct image
\[
    \begin{tikzcd}[ampersand replacement=\&]
        S \& {\{f(s) : s \in S\}} \\
        X \& Y
        \arrow[dashed, from=1-1, to=1-2]
        \arrow["\refines"{marking, allow upside down}, draw=none, from=1-1, to=2-1]
        \arrow["\refines"{marking, allow upside down}, draw=none, from=1-2, to=2-2]
        \arrow["f", from=2-1, to=2-2]
    \end{tikzcd}
\]
whose "pullbacks" are given by the inverse images
\[
    \begin{tikzcd}[ampersand replacement=\&]
        {\{x \in X \mid f(x) \in T\}} \& T \\
        X \& Y
        \arrow[dashed, from=1-1, to=1-2]
        \arrow["\refines"{marking, allow upside down}, draw=none, from=1-1, to=2-1]
        \arrow["\refines"{marking, allow upside down}, draw=none, from=1-2, to=2-2]
        \arrow["f", from=2-1, to=2-2]
    \end{tikzcd}
    \qquad\qquad\quad\
\]
and such that the internal hom of $(X, S)$ and $(Y, T)$ in $\SubSet$ is
\[
    (X \To Y
    \ ,\
    \{f : X \to Y \mid \text{for all $s \in S$, we have $f(s) \in T$}\})
\]

\begin{definition}
    \ZAP\label[definition]{def:metapull}
    The ""change of base"" of a functor $p : \E \to \B$ along a functor $F : \Ccategory \to \B$ is the functor~$\metapull{p}{F} : \metapull{\E}{F} \to \Ccategory$ defined as the following pullback of categories
    \[
        \begin{tikzcd}[ampersand replacement=\&]
            {\intro*\metapull{\E}{F}} \& \E \\
            \Ccategory\& \B
            \arrow["{\reintro*\metapull{F}{p}}", from=1-1, to=1-2]
            \arrow["{\reintro*\metapull{p}{F}}"', from=1-1, to=2-1]
            \arrow["\lrcorner"{anchor=center, pos=0.125}, draw=none, from=1-1, to=2-2]
            \arrow["p", from=1-2, to=2-2]
            \arrow["F"', from=2-1, to=2-2]
        \end{tikzcd}
    \]
    If $p$ is a "fibration"/"opfibration"/"bifibration", then so is $\metapull{p}{F}$.
\end{definition}

From an indexed viewpoint, a "change of base" along $F$ amounts to a precomposition by $F$.
The associated fibered type system obtained by "change of base" is such that
we have the reversible correspondence
\[
    \begin{prooftree}
        \hypo{ \judgein{(C, U)}{f}{(D, V)}{\metapull{p}{F}} }
        \infer[double]1{ \judgein{U}{F(f)}{V}{p} }
    \end{prooftree}
\]

We now state the following theorem,
which is a consequence of Corollary~3.8.10~and~Proposition~1.4.7 of the PhD of Hermida \cite{Hermida1993}
and of Proposition~3.2 of the work of Hasegawa \cite{hasegawa:glueing-cll}.
We prove it here using the notations of type refinement systems.

\begin{theorem}
    \ZAP\label{thm:pullback-ccc-fibration}
    For any "cartesian closed fibration"~$p : \E \to \B$
    and "cartesian functor"~$F : \Ccategory\to \B$,
    the category~$\metapull{\E}{F}$ which is the pullback
    \[
        \begin{tikzcd}[ampersand replacement=\&]
            {\metapull{\E}{F}} \& \E \\
            \Ccategory\& \B
            \arrow[from=1-1, to=1-2]
            \arrow["{\metapull{p}{F}}"', from=1-1, to=2-1]
            \arrow["\lrcorner"{anchor=center, pos=0.125}, draw=none, from=1-1, to=2-2]
            \arrow["p", from=1-2, to=2-2]
            \arrow["F"', from=2-1, to=2-2]
        \end{tikzcd}
    \]
    is "cartesian closed category", with exponentials computed as
    \[
        (C, U) \To (D, V)
        \quad:=\quad
        \left( C \To D \,,\, \pull{\ap}(U \To V) \right)
    \]
    where the canonical morphism
    \[
        \ap
        \quad:\quad
        F(C \To D)
        \ \longto\
        F(C) \To F(D)
    \]
    is  obtained by currying the image of the evaluation morphism by $F$.
    Moreover, the functor~$\metapull{p}{F} : \metapull{\E}{F} \to \Ccategory$ is "cartesian closed".
\end{theorem}

\begin{proof}
    The fact that the pullback~$F^*(\E)$ is a "cartesian category"
    and $F^*(p) : F^*(\E) \to \Ccategory$ is a "cartesian functor"
    comes from the fact that $\Ccategory$, $\B$ and $\E$ are all "cartesian categories"
    and $F : \Ccategory\to \B$ and $p : \E \to \B$ are "cartesian functors".

    Let $(C, U)$, $(D, V)$ and $(E, W)$ be three objects of $\metapull{\E}{F}$,
    and $f : E \times C \to D$ be a morphism of $\Ccategory$.
    Using the notations of type refinement systems, we establish that
    $\metapull{\E}{F}$ is a "cartesian closed category" and $\metapull{p}{F}$ is a "cartesian closed functor",
    with the internal hom written in the statement:
    \[
        \begin{prooftree}
            \hypo{ \judgein{(E, W)}{\lambda f}{(C \To D, \pull{\ap}(U \To V))}{\metapull{p}{F}} }
            \infer[double]1{ \judgein{W}{F(\lambda f)}{\pull{\ap}(U \To V)}{p} }
            \infer[double]1{ \judgein{W}{\ap \circ F(\lambda f)}{U \To V}{p} }
            \infer[double]1{ \judgein{W \times U}{F(f)}{V}{p} }
            \infer[double]1{ \judgein{W \times U}{f}{V}{\metapull{p}{F}} }
        \end{prooftree}
    \]
    where, in the penultimate step, we use the fact that
    \begin{align*}
        {\ap} \circ F(\lambda f)
        \quad & =\quad
        {\lambda F(\ev)} \circ F(\lambda f)
        \\& =\quad
        \lambda \left(F(\ev) \circ (F(\lambda f) \times \Id)\right)
        \\& =\quad
        \lambda \left(F({\ev} \circ {(\lambda f \times \Id)})\right)
        \\&=\quad
        \lambda F(f)
    \end{align*}
    This proves that $\metapull{\E}{F}$ is a "cartesian closed category" and $\metapull{p}{F}$ is a "cartesian closed functor".
\end{proof}

\section{The cartesian-closed bifibration of $Q$-regular languages}\label{sec:a-bifibration-of-Q-regular}
We recall from the introduction that $\intro*\Lam$ denotes the free cartesian closed category
generated by one object $\tyo$.
Its objects are simple types on the base type~$\tyo$,
and its morphisms are the simply-typed $\lambda$-terms modulo $\beta\eta$-conversion.
\begin{definition}
    \ZAP\label[definition]{def:sublam}
    We write $\SubLam \to \Lam$ for the "fibration"
    obtained by "change of base"
    \[
        \begin{tikzcd}[ampersand replacement=\&]
            {\intro*\SubLam} \& \SubSet \\
            \Lam \& \Set
            \arrow[from=1-1, to=1-2]
            \arrow[from=1-1, to=2-1]
            \arrow["\lrcorner"{anchor=center, pos=0.125}, draw=none, from=1-1, to=2-2]
            \arrow[from=1-2, to=2-2]
            \arrow["\Tm"', from=2-1, to=2-2]
        \end{tikzcd}
    \]
\end{definition}

More concretely, the category $\SubLam$ has as objects the pairs
\[
    (A, L)\quad\text{where $L \subseteq \Tm(A)$}
\]
and as morphisms from $(A, L)$ to $(B, K)$ the $\lambda$-terms $M : A \To B$
such that for any $\lambda$-term $N \in L$, we have $(M\ N) \in K$.

\medbreak

By \Cref{thm:pullback-ccc-fibration},
the category~$\SubLam$ is "cartesian closed"
and the functor $\SubLam \to \Lam$ is a "cartesian closed bifibration".
Note that the objects of this "bifibration" above a given type~$A$
are precisely the higher-order languages~$L$ of type~$A$,
with no assumption of regularity at this stage.

\medbreak

Now, given an object~$\Qstates$ in a cartesian closed category~$\Scategory$ such as $\Scategory=\FinSet$,
we construct a "bifibration"~$\Rec{\Qstates} \to \Lam$ of higher-order languages recognized
by the object~$\Qstates$ in the category~$\Scategory$.
We write
$$
    \begin{tikzcd}
        {\sem{-}{\Qstates}} \quad : \quad \Lam \arrow[rr] && \Scategory
    \end{tikzcd}
$$
for the unique functor (up to iso) preserving the cartesian closed structure
and determined by the equation $\sem{\tyo}{\Qstates}=\Qstates$
in the sense that it makes the diagram below commute:
\[
    \begin{tikzcd}[ampersand replacement=\&]
        \Lam \\
        1 \& \Ccategory
        \arrow["{\sem{-}{\Qstates}}", from=1-1, to=2-2]
        \arrow["\tyo", from=2-1, to=1-1]
        \arrow["\Qstates"', from=2-1, to=2-2]
    \end{tikzcd}
\]
For any type $A$, we write
$$\defTm{\Qstates}(A) \, = \,\Tm(A)/\sim_{\Qstates}$$
for the quotient of $\Tm(A)$ modulo the equivalence relation~$\sim_{\Qstates}$
induced by the semantic interpretation, defined as:
$$
    M \sim_{\Qstates} N \quad \iff \quad \sem{M}{\Qstates} = \sem{N}{\Qstates}.
$$
%

\medbreak

We are ready now to introduce the "bifibration" of higher-order languages
which are "recognized" by a given object~$\Qstates$ in a "cartesian closed category" $\Scategory$
such as $\Scategory=\FinSet$.

\begin{definition}
    \ZAP\label[definition]{def:regQ}
    Let $\Scategory$ be any "cartesian closed category" and $\Qstates$ be an object of $\Scategory$.
    We write $\Rec{\Qstates} \to \Lam$ for the "bifibration"
    obtained by the pullback of categories, or "change of base":
    \[
        \begin{tikzcd}[ampersand replacement=\&]
            {\intro*\Rec{\Qstates}} \& \SubSet \\
            \Lam \& \Set
            \arrow[from=1-1, to=1-2]
            \arrow[from=1-1, to=2-1]
            \arrow["\lrcorner"{anchor=center, pos=0.125}, draw=none, from=1-1, to=2-2]
            \arrow[from=1-2, to=2-2]
            \arrow["{\defTm{\Qstates}}"', from=2-1, to=2-2]
        \end{tikzcd}
    \]
\end{definition}
\noindent
The category $\Rec{\Qstates}$ has as objects the pairs
\[
    (A, R)\quad\text{where $R \subseteq \defTm{\Qstates}(A)$}
\]
which correspond to languages $L \subseteq \Tm(A)$ which are $\sim_{\Qstates}$-saturated,
in the sense that for all $M, N \in \Tm(A)$,
\[
    \text{if $M \in L$ and $\sem{M}{\Qstates} = \sem{N}{\Qstates}$ ,}
    \quad
    \text{then $N \in L$}
\]
By \Cref{thm:pullback-ccc-fibration}, for every object~$\Qstates$ in $\Scategory$,

\begin{proposition}\label{prop:RecQtoLam}
    The functor $\Rec{\Qstates} \to \Lam$ defines a "cartesian closed bifibration"
    of $Q$-regular languages.
\end{proposition}

\medbreak

We find it instructive to show how membership tests
can be encoded as pullbacks in the fibration~$\Rec{\Qstates} \to \Lam$.
As the type $1$ has a unique closed $\lambda$-term,
its fiber is the poset of Booleans
\[
    \Rec{\Qstates}(1)
    \quad\cong\quad
    2
    \ :=\
    \{0 < 1\}
\]
consisting in the empty and full languages.
For any $\sim_{\Qstates}$-saturated language $L\subseteq \defTm{\Qstates}(A)$
and $\lambda$-term $M$ of type~$A$ seen as a morphism $M \in \Lam(1, A)$,
the pullback
\[
    \begin{tikzcd}[ampersand replacement=\&]
        {\pull{M}(L)} \& L \\
        1 \& A
        \arrow[dashed, from=1-1, to=1-2]
        \arrow["\refines"{marking, allow upside down}, draw=none, from=1-1, to=2-1]
        \arrow["\refines"{marking, allow upside down}, draw=none, from=1-2, to=2-2]
        \arrow["M", from=2-1, to=2-2]
    \end{tikzcd}
\]
yields the Boolean value in $2=\Rec{\Qstates}(1)$
expressing whether the $\lambda$-term~$M$ belongs to the language defined~$L$.

\section{Main theorem: the cartesian-closed fibration of regular languages}
\ZAP\label{sec:fibered-adjunctions}

In \Cref{thm:pullback-nattrans},
we present a new construction which relates different "changes of base" of a common fibration,
and give in \Cref{thm:reflective-change-of-base} sufficient conditions to obtain a cartesian closed functor between the derived "fibrations".

\begin{theorem}
    \ZAP\label{thm:pullback-nattrans}
    Let $p : \E \to \B$ be a "bifibration",
    $F, G : \Ccategory\to \B$ be two functors,
    and~$\alpha : F \To G$ be a natural transformation between them.
    There are two functors
    \begin{align*}
        \pull{\alpha}
        \quad & :\quad
        \metapull{\E}{G}
        \ \longto\
        \metapull{\E}{F}
        \\
        \push{\alpha}
        \quad & :\quad
        \metapull{\E}{F}
        \ \longto\
        \metapull{\E}{G}
    \end{align*}
    which are adjoints, and which commute with the "bifibrations" into $\Ccategory$ obtained by "change of base" as follows:
    \[
        \begin{tikzcd}[ampersand replacement=\&]
            {\metapull{\E}{G}} \\
            \& {\metapull{\E}{F}} \&\& \E \\
            \& \Ccategory\&\& \B
            \arrow[""{name=0, anchor=center, inner sep=0}, "{\pull{\alpha}}", shift left=2, from=1-1, to=2-2]
            \arrow["{\metapull{G}{p}}", curve={height=-30pt}, from=1-1, to=2-4]
            \arrow["{\metapull{p}{G}}"', curve={height=30pt}, from=1-1, to=3-2]
            \arrow[""{name=1, anchor=center, inner sep=0}, "{\push{\alpha}}"{pos=0.1}, shift left=2, from=2-2, to=1-1]
            \arrow["{\metapull{F}{p}}", from=2-2, to=2-4]
            \arrow["{\metapull{p}{F}}"', from=2-2, to=3-2]
            \arrow["p", from=2-4, to=3-4]
            \arrow[""{name=2, anchor=center, inner sep=0}, "G"', curve={height=12pt}, from=3-2, to=3-4]
            \arrow[""{name=3, anchor=center, inner sep=0}, "F", curve={height=-12pt}, from=3-2, to=3-4]
            \arrow["\dashv"{anchor=center, rotate=64}, draw=none, from=1, to=0]
            \arrow["\alpha", shorten <=3pt, shorten >=3pt, Rightarrow, from=3, to=2]
        \end{tikzcd}
    \]
\end{theorem}

\begin{proof}
    We define the functors $\pull{\alpha}$ and $\push{\alpha}$ as
    \begin{align*}
        \pull{\alpha}(C, V)
        \  & :=\
        (C, \pull{\alpha_{C}}(V))
        \\
        \push{\alpha}(C, U)
        \  & :=\
        (C, \push{\alpha_{C}}(U))
    \end{align*}
    where $(C, U)$ is an object of $\metapull{\E}{F}$
    and $(C, V)$ is an object of $\metapull{\E}{G}$.
    Using the notations from type refinement systems,
    we get that for any morphism~$f : C \to D$ of $\Ccategory$,
    \[
        \begin{prooftree}
            \hypo{\judgein{(C, U)}{f}{(D, \pull{\alpha_{D}} V)}{\metapull{p}{F}}}
            \infer[double]1{\judgein{U}{F(f)}{\pull{\alpha_{D}} V}{p}}
            \infer[double]1{\judgein{U}{F(f) ; \alpha_{D}}{V}{p}}
            \infer[double]1{\judgein{U}{\alpha_{C} ; G(f)}{V}{p}}
            \infer[double]1{\judgein{\push{\alpha_{C}} U}{G(f)}{V}{p}}
            \infer[double]1{\judgein{\push{\alpha}(C, U)}{f}{(D, V)}{\metapull{p}{G}}}
        \end{prooftree}
    \]
    This establishes that we have an adjunction~$\push{\alpha} \dashv \pull{\alpha}$.
\end{proof}


\medbreak

Now, suppose given an object $\Qstates$ of a "cartesian closed category" $\Scategory$.
The object~$\Qstates$ induces a natural transformation
\[
    \begin{tikzcd}[ampersand replacement=\&]
        \Lam \&\& \Set
        \arrow[""{name=0, anchor=center, inner sep=0}, "{\defTm{\Qstates}}"', curve={height=12pt}, from=1-1, to=1-3]
        \arrow[""{name=1, anchor=center, inner sep=0}, "\Tm", curve={height=-12pt}, from=1-1, to=1-3]
        \arrow["{\sem{-}{\Qstates}}", shorten <=3pt, shorten >=3pt, Rightarrow, from=1, to=0]
    \end{tikzcd}
\]
defined as a family of functions
\begin{center}
    \begin{tikzcd}
        \sem{A}{\Qstates} \quad : \quad \Tm(A)\arrow[rr] && \defTm{\Qstates}(A)
    \end{tikzcd}
\end{center}
which sends every $\lambda$-term~$M \in \Tm(A)$
to its equivalence class modulo~$\sim_{Q}$
in the quotient $\defTm{\Qstates}(A)=\Tm(A)/\sim_{Q}$.
By \Cref{thm:pullback-nattrans}, the natural transformation
induces an adjunction which reads as follows:
\[
    \pull{\sem{-}{\Qstates}}(A, R)
    \quad=\quad
    (A, \{M \in \Tm(A) \mid \sem{M}{\Qstates} \in R\})
\]
which corresponds to the language of $\lambda$-terms obtained from a set of recognizers,
while
\[
    \push{\sem{-}{\Qstates}}(A, L)
    \quad=\quad
    (A, \{\sem{M}{\Qstates} : M \in L\})
\]
which corresponds to the smallest set of $Q$-recognizers whose language contains $L$.
These two functors assemble into the adjunction
\[
    \begin{tikzcd}[ampersand replacement=\&]
        {\Rec{X}} \&\& \SubLam \\
        \& \Lam
        \arrow[""{name=0, anchor=center, inner sep=0}, "{\pull{\sem{-}{\Qstates}}}", shift left=2, from=1-1, to=1-3]
        \arrow[curve={height=12pt}, from=1-1, to=2-2]
        \arrow[""{name=1, anchor=center, inner sep=0}, "{\push{\sem{-}{\Qstates}}}", shift left=2, from=1-3, to=1-1]
        \arrow[curve={height=-12pt}, from=1-3, to=2-2]
        \arrow["\dashv"{anchor=center, rotate=90}, draw=none, from=1, to=0]
    \end{tikzcd}
\]

We now recall the existence of a natural transformation between set-theoretic interpretations of the $\lambda$-calculus,
proven using a logical relation argument in \cite[Proposition 5.5.2]{moreau:tel-05428993}.

\begin{definition}
    \ZAP\label[definition]{def:pisem}
    Let $Q, Q'$ be two finite sets such that $|Q'| \ge |Q|$.
    For any type $A$ and $\lambda$-terms $M, N \in \Tm(A)$,
    \[
        \text{if } \sem{M}{Q'} = \sem{N}{Q'}
        \ ,\quad
        \text{then } \sem{M}{Q} = \sem{N}{Q}
    \]
    This defines a natural transformation
    \[
        \begin{tikzcd}[ampersand replacement=\&]
            \Lam \&\& \Set
            \arrow[""{name=0, anchor=center, inner sep=0}, "{\defTm{Q}}"', curve={height=12pt}, from=1-1, to=1-3]
            \arrow[""{name=1, anchor=center, inner sep=0}, "{\defTm{Q'}}", curve={height=-12pt}, from=1-1, to=1-3]
            \arrow["\pisem", shorten <=3pt, shorten >=3pt, Rightarrow, from=1, to=0]
        \end{tikzcd}
    \]
    whose components are surjective.
\end{definition}

By \Cref{thm:pullback-nattrans},
the natural transformation $\pisem$ of \Cref{def:pisem} induces an adjunction
\[
    \begin{tikzcd}[ampersand replacement=\&]
        {\Rec{Q'}} \&\& {\Rec{Q}} \\
        \& \Lam
        \arrow[""{name=0, anchor=center, inner sep=0}, "{\push{\pisem}}"', shift right=2, from=1-1, to=1-3]
        \arrow[curve={height=12pt}, from=1-1, to=2-2]
        \arrow[""{name=1, anchor=center, inner sep=0}, "{\pull{\pisem}}"', shift right=2, from=1-3, to=1-1]
        \arrow[curve={height=-12pt}, from=1-3, to=2-2]
        \arrow["\dashv"{anchor=center, rotate=90}, draw=none, from=0, to=1]
    \end{tikzcd}
\]
such that $\pull{\pi}$ includes the $Q$-recognizable languages into the $Q'$-recognizable ones,
and $\push{\pi}$ performs a $Q$-saturation.
By taking the colimit of the chain made of all the inclusions $\pull{\pi}$,
we recover in this way all the higher-order regular languages.

\begin{definition}
    \ZAP\label[definition]{def:reg-colimit}
    We consider the "fibration" defined as the colimit
    \[
        \begin{tikzcd}[ampersand replacement=\&]
            {\intro*\Reg} \\
            \Lam
            \arrow[from=1-1, to=2-1]
        \end{tikzcd}
        \quad:=\quad
        \colim\
        \begin{tikzcd}[ampersand replacement=\&]
            {\Rec{[1]}} \& {\Rec{[2]}} \& \dots \\
            \& \Lam
            \arrow["{\pull{\pisem}}", from=1-1, to=1-2]
            \arrow[from=1-1, to=2-2]
            \arrow[from=1-2, to=1-3]
            \arrow[from=1-2, to=2-2]
            \arrow[from=1-3, to=2-2]
        \end{tikzcd}
    \]
    in the category of fibrations over~$\Lam$.
\end{definition}

We are now interested in knowing when the right adjoint obtained from \Cref{thm:pullback-nattrans}
\[
    \pull{\alpha}
    \quad:\quad
    \metapull{E}{G}
    \ \longto\
    \metapull{E}{F}
\]
which already preserves cartesian products,
also preserves the internal hom and thus is a cartesian closed functor.
To give sufficient conditions for this, we introduce the following notion.

\begin{definition}
    \ZAP\label[definition]{def:reflective-morphism}
    Let $p : \E \to \B$ be a "bifibration".
    A morphism~$f : X \to Y$ of $\B$ is said to be ""reflective"" when the adjunction
    \[
        \begin{tikzcd}[ampersand replacement=\&]
            {\E_{X}} \&\& {\E_{Y}}
            \arrow[""{name=0, anchor=center, inner sep=0}, "{\push{f}}"', shift right=2, from=1-1, to=1-3]
            \arrow[""{name=1, anchor=center, inner sep=0}, "{\pull{f}}"', shift right=2, from=1-3, to=1-1]
            \arrow["\dashv"{anchor=center, rotate=90}, draw=none, from=0, to=1]
        \end{tikzcd}
    \]
    is reflective, i.e. when the counit obtained by the derivation
    \[
        \begin{prooftree}
            \infer0{\judge{\pull{f} U}{\Id_{X}}{\pull{f} U}}
            \infer1{\judge{\pull{f} U}{f}{U}}
            \infer1{\judge{\push{f} \pull{f} U}{\Id_{Y}}{U}}
        \end{prooftree}
    \]
    is an isomorphism for every $U \in \E_{Y}$.
\end{definition}

\begin{example}
    The "reflective morphisms" of the "bifibration" $\SubSet \to \Set$ are the set-theoretic functions~$f : X \to Y$ such that
    \[
        \begin{tikzcd}[ampersand replacement=\&]
            {\wp(Y)} \& {\wp(X)} \& {\wp(Y)}
            \arrow["{f^{-1}}", from=1-1, to=1-2]
            \arrow["f", from=1-2, to=1-3]
        \end{tikzcd}
    \]
    is the identity, i.e. $f$ is a surjective function.

    \medbreak

    A categorified version of the previous example is given by the "bifibration" $\PSh \to \Cat$ \cite[\S3.1]{MelliesZeilberger18},
    which offers a fibered view on the presheaf hyperdoctrine without Beck-Chevalley condition \cite{LawvereEqHyp}.
    In this "bifibration", the "pullbacks" are computed by precomposition
    and the "pushforwards" are computed by left Kan extensions.
    For any categories $\Xcategory$ and $\Ycategory$,
    the functors~$F : \Xcategory \to \Ycategory$ such that
    \[
        \begin{tikzcd}[ampersand replacement=\&]
            {\PSh(\Ycategory)} \&\& {\PSh(\Xcategory)} \&\& {\PSh(\Ycategory)}
            \arrow["{(-) \circ F}", from=1-1, to=1-3]
            \arrow["{\Lan_F}", from=1-3, to=1-5]
        \end{tikzcd}
    \]
    is naturally isomorphic to the identity on $\PSh(\Ycategory)$,
    i.e. the "reflective morphisms" of this "bifibration",
    are known as lax epimorphisms \cite{laxepimorphisms}.
\end{example}

We introduced in the notion of "reflective morphism" in \Cref{def:reflective-morphism}
as it allows to formulate a sufficient condition for the right adjoint obtained from \Cref{thm:pullback-nattrans} to preserve the internal hom.
This is the content of the following theorem, whose proof uses the Frobenius reciprocity principle.

\begin{theorem}
    \ZAP\label{thm:reflective-change-of-base}
    For any "cartesian closed bifibration"~$p$,
    two "cartesian functors"~$F$ and $G$,
    and a natural transformation $\theta : F \To G$,
    i.e. a diagram of the shape
    \[
        \begin{tikzcd}[ampersand replacement=\&]
            \&\& \E \\
            \Ccategory\&\& \B
            \arrow["p", from=1-3, to=2-3]
            \arrow[""{name=0, anchor=center, inner sep=0}, "G"', curve={height=12pt}, from=2-1, to=2-3]
            \arrow[""{name=1, anchor=center, inner sep=0}, "F", curve={height=-12pt}, from=2-1, to=2-3]
            \arrow["\theta", shorten <=3pt, shorten >=3pt, Rightarrow, from=1, to=0]
        \end{tikzcd}
    \]
    If all the components of the natural transformation~$\theta$ are "reflective morphisms" for the "bifibration"~$p$,
    then the functor
    \[
        \pull{\theta}
        \quad:\quad
        \metapull{\E}{G}
        \ \longto\
        \metapull{\E}{F}
    \]
    is a "cartesian closed functor".
\end{theorem}

\begin{proof}
    By definition of "reflective morphisms",
    for any object~$(D, V)$ of the category~$\metapull{\E}{G}$ we have that
    \[
        \push{\alpha} \pull{\alpha} (D, V)
        \quad=\quad
        (D, \push{\alpha_{D}} \pull{\alpha_{D}} V)
        \quad\cong\quad
        (D, V)
    \]
    where this morphism is the counit of the adjunction.
    Therefore, we obtain that the adjunction
    \[
        \begin{tikzcd}[ampersand replacement=\&]
            {\metapull{\E}{G}} \&\& {\metapull{\E}{F}}
            \arrow[""{name=0, anchor=center, inner sep=0}, "{\pull{\alpha}}", shift left=2, from=1-1, to=1-3]
            \arrow[""{name=1, anchor=center, inner sep=0}, "{\push{\alpha}}", shift left=2, from=1-3, to=1-1]
            \arrow["\dashv"{anchor=center, rotate=90}, draw=none, from=1, to=0]
        \end{tikzcd}
    \]
    is reflective.
    As the functor~$\push{\alpha}$ preserves cartesian products,
    Frobenius reciprocity \cite[A4.3.1]{10.1093/oso/9780198515982.001.0001}
    gets us that $\pull{\alpha}$ is a "cartesian closed functor".
\end{proof}

As a consequence of \Cref{thm:reflective-change-of-base},
all the right adjoint functors
\[
    \pull{\pisem}
    \quad:\quad
    \Rec{Q}
    \ \longto\
    \Rec{Q'}
    \qquad\text{for all $|Q'| \ge |Q|$}
\]
are cartesian closed,
and thus their filtered colimit is a cartesian closed category.
Indeed, the internal hom $S \To T$ is obtained in the colimit category
by taking a large enough $Q'$ so both objects $S$ and $T$ belong to the same cartesian closed category~$\Rec{Q'}$, where one computes the internal hom.

\begin{theorem}[Thm C.]
    \ZAP\label{thm:regcolim-ccc-fibration}
    The functor $p:\Reg \to \Lam$ is a "cartesian closed fibration".
\end{theorem}

\section{A classifying fibration of clopen predicates over Stone spaces}
\ZAP\label{sec:fibration-topological-predicates}

Our approach until now
consisted in building multiple fibrations by change of base
solely from the fibration $\SubSet \to \Set$.
This fibration has a universal property:
it classifies fibrations whose fibers are complete atomic Boolean algebras
and whose reindexing functors behave accordingly.
Our object of interest however,
namely the fibration $\Reg \to \Lam$ built in \Cref{def:reg-colimit} as colimit,
cannot be obtained as change of base from $\SubSet \to \Set$,
for its fibers are Boolean algebras that are atomic, but not complete in general.

For this reason,
we showcase another construction which amounts to replace the fibration $\SubSet \to \Set$ by another one of a more topological flavor,
inspired by Stone duality.

\begin{definition}
    \ZAP\label[definition]{def:clopencat}
    We write $\intro*\Clopen$ for the category
    \begin{itemize}
        \item whose objects are pairs~$(X, S)$ of a Stone space~$X$ together with a ""clopen"" subset~$S \subseteq X$, i.e. a set of points which is both open and closed for the topology carried by the space~$X$
        \item whose morphisms~$(X, S) \to (Y, T)$ are continuous functions~$f : X \to Y$ such that for any $x \in X$,
              \[
                  \text{if}\
                  x \in S
                  \ ,
                  \quad\text{then}\
                  f(x) \in T
              \]
    \end{itemize}
    The functor~$(X, S) \mapsto X$ is a "fibration"~$\Clopen \to \Stone$.
\end{definition}

This fibration can equivalently be constructed in various ways.
One consists in the Grothendieck construction
applied to the clopen functor $\Stone \to \op{\BA}$
at the heart of Stone duality.
Another one consists in the externalization of the usual poset of Booleans $2 := \{0 < 1\}$
seen as internal to $\FinSet$, and hence to $\Stone$.
Indeed, objects~$(X, S)$ of $\Clopen$ correspond to morphisms~$S \in \Stone(X, 2)$,
and morphisms $f : (X, S) \to (Y, T)$ of $\Clopen$ correspond to factorizations
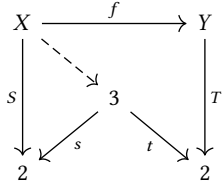
\[
    \begin{tikzcd}[ampersand replacement=\&]
        X \&\& Y \\
        \& 3 \\
        2 \&\& 2
        \arrow["f", from=1-1, to=1-3]
        \arrow[dashed, from=1-1, to=2-2]
        \arrow["S"', from=1-1, to=3-1]
        \arrow["T", from=1-3, to=3-3]
        \arrow["s", from=2-2, to=3-1]
        \arrow["t"', from=2-2, to=3-3]
    \end{tikzcd}
\]
through the discrete space $3 := \{(0, 0), (0, 1), (1, 1)\}$ of inequalities in the poset $2$,
with the projections $s(b, b') := b$ and $t(b, b') := b'$.

We now show that $\Clopen \to \Stone$ has a classifying property,
with respect to fibrations of Boolean algebras.

\begin{theorem}
    \label{thm:clopen-classifies-ba}
    If a fibration $p : \E \to \B$ is such that
    its fibers are Boolean algebras
    and its reindexing functors are Boolean algebra homomorphisms,
    then it can be obtained as a change of base as follows:
    
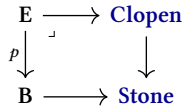
\[
        \begin{tikzcd}[ampersand replacement=\&]
            \E \& \Clopen \\
            \B \& \Stone
            \arrow[from=1-1, to=1-2]
            \arrow["p"', from=1-1, to=2-1]
            \arrow["\lrcorner"{anchor=center, pos=0.125}, draw=none, from=1-1, to=2-2]
            \arrow[from=1-2, to=2-2]
            \arrow[from=2-1, to=2-2]
        \end{tikzcd}
    \]
\end{theorem}

\begin{proof}
    \AP We write $\intro*\BA$ for the category of Boolean algebras and their homomorphisms,
    and $\intro*\BApt$ for the category whose objects are Boolean algebras paired with one of their element $(B, b)$,
    and whose morphisms $(B, b) \to (B', b')$ are homomorphisms $u$ such that $u(b) \le b'$.
    The hypothesis on $p$ corresponds exactly to the fact that it can be obtained as a "change of base"
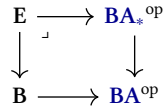
    \[
        \begin{tikzcd}[ampersand replacement=\&]
            \E \& {\op{\BApt}} \\
            \B \& {\op{\BA}}
            \arrow[from=1-1, to=1-2]
            \arrow[from=1-1, to=2-1]
            \arrow["\lrcorner"{anchor=center, pos=0.125}, draw=none, from=1-1, to=2-2]
            \arrow[from=1-2, to=2-2]
            \arrow[from=2-1, to=2-2]
        \end{tikzcd}
    \]
    Stone duality informs us that we have an adjoint equivalence
    \[
        \begin{tikzcd}[ampersand replacement=\&]
            {\op{\BA}} \& \Stone
            \arrow[""{name=0, anchor=center, inner sep=0}, shift left=2, from=1-1, to=1-2]
            \arrow[""{name=1, anchor=center, inner sep=0}, shift left=2, from=1-2, to=1-1]
            \arrow["\dashv"{anchor=center, rotate=-90}, draw=none, from=0, to=1]
        \end{tikzcd}
    \]
    which sends a Boolean algebra on its Stone space of ultrafilters,
    and a Stone space on its Boolean algebra of clopen subsets.
    We observe that this adjoint equivalence lifts to an adjoint equivalence of fibrations
    
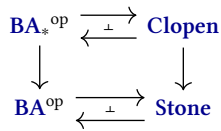
\[
        \begin{tikzcd}[ampersand replacement=\&]
            {\op{\BA_*}} \& \Clopen \\
            {\op{\BA}} \& \Stone
            \arrow[""{name=0, anchor=center, inner sep=0}, shift left=2, from=1-1, to=1-2]
            \arrow[from=1-1, to=2-1]
            \arrow[""{name=1, anchor=center, inner sep=0}, shift left=2, from=1-2, to=1-1]
            \arrow[from=1-2, to=2-2]
            \arrow[""{name=2, anchor=center, inner sep=0}, shift left=2, from=2-1, to=2-2]
            \arrow[""{name=3, anchor=center, inner sep=0}, shift left=2, from=2-2, to=2-1]
            \arrow["\dashv"{anchor=center, rotate=-90}, draw=none, from=0, to=1]
            \arrow["\dashv"{anchor=center, rotate=-90}, draw=none, from=2, to=3]
        \end{tikzcd}
    \]
    hence we get a pullback square
    
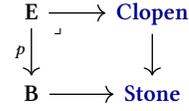
\[
        \begin{tikzcd}[ampersand replacement=\&]
            \E \& \Clopen \\
            \B \& \Stone
            \arrow[from=1-1, to=1-2]
            \arrow["p"', from=1-1, to=2-1]
            \arrow["\lrcorner"{anchor=center, pos=0.125}, draw=none, from=1-1, to=2-2]
            \arrow[from=1-2, to=2-2]
            \arrow[from=2-1, to=2-2]
        \end{tikzcd}
    \]
    whose bottom functor sends an object $B$ of $\B$ to the space of ultrafilters on of the Boolean algebra of predicates over $B$.
\end{proof}

Similarly, the fibration $\SubSet \to \Set$ has a classifying property with respect to complete atomic Boolean algebras.

Given that an ultrafilter is a type in the model-theoretic sense,
hence a model,
we think of the functor $\B \to \Stone$ built in \Cref{thm:clopen-classifies-ba}
as associating to any object $B$ of $\B$ a canonical space of models of the logic of predicates over $B$.

Now that we defined our topological fibration and described its classifying property,
we are interested in knowing which morphisms have pushforwards.
This question has been studied in the literature on polyadic spaces \cite[Proposition~4.7]{vanGool2024},
and the morphisms having pushforwards have been characterized as follows.

\begin{proposition}
    \label[proposition]{prop:open}
    For any morphism $f \in \Stone(X, Y)$,
    the following statements are equivalent:
    \begin{enumerate}
        \item the morphism $f$ has pushforwards
        \item direct images by $f$ preserve clopen subsets
        \item direct images by $f$ preserve open subsets
    \end{enumerate}
    If these equivalent conditions hold, we say that $f$ is open.
\end{proposition}




Given that all complete atomic Boolean algebras can be seen as mere Boolean algebras,
we can apply \Cref{thm:clopen-classifies-ba} to the fibration $\SubSet \to \Set$ itself.
This gets us the pullback square
\[
    \begin{tikzcd}[ampersand replacement=\&]
        \SubSet \& \Clopen \\
        \Set \& \Stone
        \arrow[from=1-1, to=1-2]
        \arrow[from=1-1, to=2-1]
        \arrow["\lrcorner"{anchor=center, pos=0.125}, draw=none, from=1-1, to=2-2]
        \arrow[from=1-2, to=2-2]
        \arrow["\beta", from=2-1, to=2-2]
    \end{tikzcd}
\]
where $\beta : \Set \to \Stone$ sends a set on the \StoneCech compactification of its associated discrete space.
Indeed, clopen subsets of $\beta X$ correspond to subsets of $X$.

We think of the fibration $\Clopen \to \Stone$ as a topological refinement of $\SubSet \to \Set$.
In the rest of this section, we demonstrate this viewpoint by showing how to deduce three key properties of the subset fibration from the clopen one.

\medskip
\noindent
\textbf{Pushforwards.}
The first one concerns the fact that $\SubSet \to \Set$ is a bifibration,
i.e. all the set-theoretic functions have pushforwards.
To relate that phenomenon to Stone spaces,
it is instructive to notice that the functor $\beta : \Set \to \Stone$
is left adjoint to the functor sending a space on its underlying set of points.
In this way, we get an adjunction between refinement systems in the sense of \cite[\S2.2]{MelliesZeilberger18}
\[
    \begin{tikzcd}[ampersand replacement=\&]
        \SubSet \& \Clopen \\
        \Set \& \Stone
        \arrow[""{name=0, anchor=center, inner sep=0}, shift left=2, from=1-1, to=1-2]
        \arrow[from=1-1, to=2-1]
        \arrow[""{name=1, anchor=center, inner sep=0}, shift left=2, from=1-2, to=1-1]
        \arrow[from=1-2, to=2-2]
        \arrow[""{name=2, anchor=center, inner sep=0}, "\beta", shift left=2, from=2-1, to=2-2]
        \arrow[""{name=3, anchor=center, inner sep=0}, "{\text{points}}", shift left=2, from=2-2, to=2-1]
        \arrow["\dashv"{anchor=center, rotate=-90}, draw=none, from=0, to=1]
        \arrow["\dashv"{anchor=center, rotate=-90}, draw=none, from=2, to=3]
    \end{tikzcd}
\]
A consequence of that fact is that for any set-theoretic functions $f \in \Set(X, Y)$,
the associated continuous function $\beta f \in \Stone(\beta X, \beta Y)$ has pushforwards, computed using the ones of $f$, as proven in all generality in \cite[Proposition 2.3]{MelliesZeilberger18}.

Therefore,
although $\Clopen \to \Stone$ is not a bifibration,
it contains all the pushforwards that we have in $\SubSet \to \Set$.

\medskip
\noindent
\textbf{The cartesian products.}
The functor $\beta : \Set \to \Stone$ does not preserve products.
Yet, as any functor between cartesian monoidal categories,
it has a unique oplax structure \cite[\S5.2]{Mellies09panorama}
which we write
\[
    n_{X, Y}
    \quad:\quad
    \beta(X \times Y)
    \ \longto\
    \beta X \times \beta Y
    \qquad\text{for any sets $X$ and $Y$}
\]
Let $U$ and $V$ be subsets of the sets $X$ and $Y$ respectively,
seen as clopen subsets of the spaces $\beta X$ and $\beta Y$.
Taking their products as objects of $\Clopen$ yields a clopen of the space $\beta X \times \beta Y$,
which can be pulled back along the oplax structure.
\[
    \begin{tikzcd}[ampersand replacement=\&]
        {\pull{n_{X, Y}}(U \times V)} \&\& {U \times V} \\
        {\beta (X \times Y)} \&\& {\beta X \times \beta Y}
        \arrow[dashed, from=1-1, to=1-3]
        \arrow["\sqsubset"{marking, allow upside down}, draw=none, from=1-1, to=2-1]
        \arrow["\sqsubset"{marking, allow upside down}, draw=none, from=1-3, to=2-3]
        \arrow["{n_{X, Y}}"', from=2-1, to=2-3]
    \end{tikzcd}
\]
In this way, we recover exactly the cartesian product of $U$ and $V$ as objects of $\SubSet$.
This shows that the cartesian structure of $\SubSet$ can be built from the one of $\Clopen$ and the oplax structure.

\medskip
\noindent
\textbf{The internal hom.}
One of the important appeal of the present theory is to generalize the usual theory of regular languages of words to the higher-order.
Therefore, the fact that $\Clopen \to \Stone$ is not a cartesian closed fibration may seem like an obstruction to our higher-order approach.
Despite this lack of closure,
the fibration $\Clopen \to \Stone$ has simple products,
a property from the theory of fibrations \cite[Definition~1.9.1]{JacobsCLTT}
which we recall for the reader's convenience.

\begin{definition}
    \ZAP\label[definition]{def:simple-products}
    A fibration $p : \E \to \B$,
    where $\B$ is a cartesian category,
    is said to have ""simple products"" if
    for any objects~$X$ and~$Y$,
    and writing~$\pi : X \times Y \to X$ for the associated left projection morphism,
    there exists a right adjoint
    \[
        \pull{\pi}
        \dashv
        \copush{\pi}
    \]
    and the following Beck-Chevalley condition holds,
    i.e. that for any morphism $u : Z \to X$ of $\B$,
    the canonical natural transformation
    \[
        \begin{tikzcd}[ampersand replacement=\&]
            {\E(X \times Y)} \&\& {\E(X)} \\
            {\E(Z \times Y)} \&\& {\E(Z)}
            \arrow["{\copush{\pi}}", from=1-1, to=1-3]
            \arrow["{\pull{f \times \Id}}"', from=1-1, to=2-1]
            \arrow[between={0.2}{0.8}, Rightarrow, from=1-3, to=2-1]
            \arrow["{\pull{f}}", from=1-3, to=2-3]
            \arrow["{\copush{\pi}}"', from=2-1, to=2-3]
        \end{tikzcd}
    \]
    is an isomorphism.
\end{definition}

An important topological property of projections
\[
    \pi
    \quad:\quad
    X \times Y
    \ \longto\
    X
\]
between Stone spaces is that they are open,
in the sense of \Cref{prop:open} \cite[Exercise~2.2.2]{Gehrke_vanGool_2024}.
As a consequence,
there exists a left adjoint
\[
    \push{\pi} \dashv \pull{\pi}
\]
and, given that fibers are Boolean algebras,
we deduce the existence of a right adjoint through the usual formula
\[
    \copush{\pi}
    \quad:=\quad
    \neg \circ \push{\pi} \circ \neg
\]
and the Beck-Chevalley condition holds as pushforwards are computed by direct images.
As a consequence, the fibration $\Clopen \to \Stone$ has simple products.

This simple fact has far-reaching consequences,
as studied by Hermida in his PhD thesis.
In particular,
we deduce that $\SubSet \to \Set$ has simple products,
using an argument close to \cite[Proposition~1.4.7]{Hermida1993}.
In turn, fibrations with simple products whose base are closed are themselves cartesian closed \cite[Corollary~3.3.10]{Hermida1993}.
This demonstrates that the cartesian closed nature of the fibration $\SubSet \to \Set$
can be deduced from the simple products of $\Clopen \to \Stone$.

\section{The cartesian-closed fibration of regular languages using the profinite $\lambda$-calculus}
\label{sec:reg-through-profinite-lambda-terms}

One starting point of the recent work on the profinite $\lambda$-calculus \cite{entics:12280}
was to apply Stone duality to higher-order regular languages in order to develop a topological approach to their recognition.
The central definition of the paper was of a space $\ProTm(A)$,
for every type $A$,
whose clopen sets are in natural correspondence with the regular languages in the Boolean algebra~$\Reg(A)$.

The collection of all these spaces forms a functor
\[
    \ProTm
    \quad:\quad
    \Lam
    \ \longto\
    \Stone
\]
and the aforementioned observation on clopens amounts to the fact that we get the fibration of higher-order regular languages by change of base along this functor:
\[
    \begin{tikzcd}[ampersand replacement=\&]
        \Reg \& \Clopen \\
        \Lam \& \Stone
        \arrow[from=1-1, to=1-2]
        \arrow[from=1-1, to=2-1]
        \arrow["\lrcorner"{anchor=center, pos=0.125}, draw=none, from=1-1, to=2-2]
        \arrow[from=1-2, to=2-2]
        \arrow["\ProTm", from=2-1, to=2-2]
    \end{tikzcd}
\]
We are therefore able to give a second definition of the fibration $\Reg \to \Lam$,
by a single change of base of the topological fibration $\Clopen \to \Stone$.

To give a topological proof that $\Reg \to \Lam$ is a "cartesian closed fibration",
we use the following theorem from Hermida's PhD thesis \cite[Corollary~3.3]{Hermida1993}.

\begin{theorem}
    \label{thm:hermida-simple-products}
    If $p : \E \to \B$ is a "fibration" with "simple products" such that
    $\B$ is cartesian closed
    and each fiber of $p$ is cartesian closed,
    then $\E$ is cartesian closed and $p$ is a "cartesian closed fibration".
\end{theorem}

Assuming the hypotheses of \Cref{thm:hermida-simple-products},
the internal hom in $\E$ of two objects $U \refines X$ and $V \refines Y$ is computed as
\[
    \copush{\pi}\left(\,\pull{\pi'}(U)\ \supset\ \pull{\ev}(V)\,\right)
    \quad\refines\quad
    X \To Y
\]
where $\pi : (X \To Y) \times X \to X \To Y$ and $\pi' : (X \To Y) \times X \to X$ are the left and right projections,
and $\supset$ is the internal hom of the fiber category~$\E((X \To Y) \times X)$.

A crucial property of the functor $\ProTm$,
proven in~\cite[Proposition~8.1.8]{moreau:tel-05428993},
is that it preserves finite products.
As a consequence,
we get that $\Reg \to \Lam$ has "simple products" from the fact that $\Clopen \to \Stone$ does \cite[Proposition~1.4.7]{Hermida1993}.
As the fibers $\Reg(A)$ are Boolean algebras for any type $A$, hence cartesian closed,
we can apply \Cref{thm:hermida-simple-products} and obtain a second, topological proof of our main theorem.

\begin{theorem}[Thm C.]
    \ZAP\label{thm:reg-stone-ccc-fibration}
    The functor $p:\Reg \to \Lam$ is a "cartesian closed fibration".
\end{theorem}

Following the discussion in \Cref{sec:fibration-topological-predicates} on the morphisms of $\Stone$ that admit pushforwards,
we introduce the following definition.

\begin{definition}
    \ZAP\label[definition]{def:open-lambda-term}
    Let $A$ and $B$ be two types.
    A $\lambda$-term~$M \in \Lam(A, B)$ is ""open"" when the associated continuous function
    \[
        \ProTm(M)
        \quad:\quad
        \ProTm(A)
        \ \longto\
        \ProTm(B)
    \]
    between the Stone spaces of profinite $\lambda$-terms, is an open function.
\end{definition}

We recall that, by \Cref{prop:open},
a $\lambda$-term $M \in \Lam(A, B)$ is open if it preserves regular languages by direct image.
We stress the fact that one cannot curry the types in \Cref{def:open-lambda-term}.
Indeed, for any $\lambda$-term~$M \in \Lam(A, B)$,
the $\lambda$-terms
\[
    \lambda M
    \ \in\
    \Lam(1, A \To B)
\]
is always "open", while $M$ itself is possibly not "open",
cf. the examples in \Cref{sec:logic-of-regular-languages}.

We finish this section by introducing another fibration,
that this turn to Stone spaces reveals as more fundamental than $\Reg \to \Lam$.
Just as types and $\lambda$-terms form a cartesian closed category $\Lam$,
types and profinite $\lambda$-terms form a $\Stone$-enriched cartesian closed category $\ProLam$,
studied in detail in~\cite[Chapter~8]{moreau:tel-05428993}.
The fact that every $\lambda$-term may be seen as a profinite $\lambda$-term takes the form of an identity-on-objects functor
\[
    \Lam
    \ \longto\
    \ProLam
\]
which is faithful by Statman's theorem \cite[Theorem~2]{DBLP:journals/jsyml/Statman82}, \cite{https://doi.org/10.48550/arxiv.2309.03602}.
We therefore decompose the functor $\ProTm$ as the composite
\[
    \begin{tikzcd}[ampersand replacement=\&]
        \Lam \& \ProLam \&\& \Stone
        \arrow[from=1-1, to=1-2]
        \arrow["{\ProLam(1, -)}", from=1-2, to=1-4]
    \end{tikzcd}
\]
which, in turn, induces a new cartesian closed fibration $\ProReg \to \ProLam$
as in the following diagram of changes of bases
\[
    \begin{tikzcd}[ampersand replacement=\&]
        \Reg \& \ProReg \&\& \Clopen \\
        \Lam \& \ProLam \&\& \Stone
        \arrow[from=1-1, to=1-2]
        \arrow[from=1-1, to=2-1]
        \arrow["\lrcorner"{anchor=center, pos=0.125}, draw=none, from=1-1, to=2-2]
        \arrow[from=1-2, to=1-4]
        \arrow[from=1-2, to=2-2]
        \arrow[from=1-4, to=2-4]
        \arrow[from=2-1, to=2-2]
        \arrow[""{name=0, anchor=center, inner sep=0}, "{\ProLam(1, -)}", from=2-2, to=2-4]
        \arrow["\lrcorner"{anchor=center, pos=0.125}, draw=none, from=1-2, to=0]
    \end{tikzcd}
\]
We also mention that \Cref{def:open-lambda-term} extends straightforwardly to a notion of open profinite $\lambda$-term.
For example, among the profinite $\lambda$-terms $M \in \ProLam(1, A)$,
the open ones are the isolated points, which are the usual $\lambda$-terms $M \in \Lam(1, A)$ which are all open by Statman's theorem.
This opens up new perspective for profinite quantifiers.

\section{A conservative extension of Brzozowski derivatives to higher-order languages}
\ZAP\label{sec:logic-of-regular-languages}

In this section, we will illustrate the benefits
of Theorem C. (Thms~\ref{thm:regcolim-ccc-fibration})
by generalizing the notion of Brzozowski to the higher-order setting of the $\lambda$-calculus,
To that purpose, we use an Isbell-like adjunction in the sense of Melliès and Zeilberger \cite{MelliesZeilberger18}.
In the traditional case of finite words,
the left and right derivatives of a language $L \subseteq \Sigma^*$ by a letter $a \in \Sigma$
are the two languages
\begin{align*}
    a \backslash L \quad & := \quad \{w \in \Sigma^* \mid aw \in L\}
    \\
    L \slash a \quad     & := \quad \{w \in \Sigma^* \mid wa \in L\}
\end{align*}
and these are contravariant adjoints one of the other.
As hinted in the introduction,
this construction generalizes to the higher-order setting as follows:

\begin{theorem}
    \label{thm:brzozowski}
    Let $A$, $B$ and $C$ be three types.
    For any $\lambda$-term
    \[
        M
        \quad:\quad
        A \times B \To C
    \]
    and regular language $L_C\in \Reg(C)$,
    we have an adjunction
    \[
        \begin{tikzcd}[ampersand replacement=\&]
            {\Reg(A)} \&\&\& {\op{\Reg(B)}}
            \arrow[""{name=0, anchor=center, inner sep=0}, "{L_A\ \mapsto\ L_A \backslash L_C}", shift left=2, from=1-1, to=1-4]
            \arrow[""{name=1, anchor=center, inner sep=0}, "{L_C \slash L_B\ \mapsfrom\ L_B}", shift left=2, from=1-4, to=1-1]
            \arrow["\dashv"{anchor=center, rotate=-90}, draw=none, from=0, to=1]
        \end{tikzcd}
    \]
    where the residuals are defined as
    \begin{align*}
        L_A \backslash L_C
        \quad := & \quad
        \pull{\lambda x_B. \lambda x_A. M\ (x_A, x_B)}(L_A \To L_C)
        \\
                 & \quad
        \big\{\,N_B : B \mid \forall N_A \in L_A, \text{ we have } M\ (N_A, N_B) \in L_C\,\big\}
        \\
        L_C \slash L_B
        \quad := & \quad
        \pull{\lambda x_A. \lambda x_B. M\ (x_A, x_B)}(L_B \To L_C)
        \\
                 & \quad
        \big\{\, N_A : A \mid \forall N_B \in L_B, \text{ we have } M\ (N_A, N_B) \in L_C\,\big\}
    \end{align*}
\end{theorem}

We get back the usual notion of Brzozowski derivative when $A = B = C = \Words{\Sigma}$, and $M$ is the $\lambda$-term
\begin{align*}
    M
    \quad & : \quad
    \Words{\Sigma} \To
    \Words{\Sigma} \To
    \Words{\Sigma}
    \\
    M
    \quad & := \quad
    \lambda u. \lambda v. \lambda \vec{a}. \lambda x. u\ \vec{a}\ (v\ \vec{a}\ x)
\end{align*}
whose action on closed $\lambda$-terms corresponds to the concatenation of words
through the Church encoding.

To illustrate further this general Brzozowski derivative,
we now proceed to show how it can be applied to the case of finite ranked trees (with sharing of variables).
To any ranked alphabet, i.e. finite first order signature
\[
    \Sigma
    \quad:=\quad
    \{a_1 : n_1, \dots, a_n : n_l\}
\]
we associate the type
\[
    \Tree{\Sigma}
    \quad:=\quad
    (\tyo^{n_1} \To \tyo) \times \dots \times (\tyo^{n_l} \To \tyo)\ \To \ \tyo
\]
whose closed $\lambda$-terms modulo $\beta\eta$-conversion correspond to finite ranked trees over $\Sigma$.
For any ranked alphabet $\Sigma$,
we define another ranked alphabet
\[
    \Sigma+1
    \quad:=\quad
    \{a_1 : n_1, \dots, a_l : n_l, b : 0\}
\]
obtained from $\Sigma$ by adding a constant $b$.
Every ranked tree on $\Sigma+1$ can be seen as a tree-contexts~$k$
waiting for a tree~$t$ on $\Sigma$ to be grafted on the constant~$b$,
resulting in another tree~$k[t]$ on $\Sigma$.
The grafting operation can be defined as the simply typed $\lambda$-term
\begin{align*}
    M \quad & :\quad \Tree{\Sigma+1} \To \Tree{\Sigma} \To \Tree{\Sigma}
    \\
    M \quad & :=\quad \lambda k. \lambda t. \lambda \vec{a}. k\ (\vec{a}\ \,,\, t\ \vec{a})
\end{align*}
Note that
$$
    M (k\, , \, t \, )
    \,\, \cong_{\beta\eta} \,\, k[t]
$$
By \Cref{thm:brzozowski},
any regular tree language $L$ induces an adjunction
\[
    \begin{tikzcd}[ampersand replacement=\&]
        {\Reg(\Tree{\Sigma+1})} \&\& {\op{\Reg(\Tree{\Sigma})}}
        \arrow[""{name=0, anchor=center, inner sep=0}, shift left=2, from=1-1, to=1-3]
        \arrow[""{name=1, anchor=center, inner sep=0}, shift left=2, from=1-3, to=1-1]
        \arrow["\dashv"{anchor=center, rotate=-90}, draw=none, from=0, to=1]
    \end{tikzcd}
\]
Suppose given a regular language of trees~$L$.
Every regular language of tree-contexts~$K$ induces the regular language of trees
\[
    K \backslash L
    \quad :=\quad
    \big\{ \,t : \Tree{\Sigma} \mid \forall k \in K, \text{we have } k[t] \in L \, \big\}
    \hspace{.25cm}
\]
and conversely, every regular language of trees~$L$ induces the regular language of tree-contexts
\[
    L \slash L'
    \quad :=\quad
    \big\{\, k : \Tree{\Sigma+1} \mid \forall t \in L', \text{we have } k[t] \in L \, \big\}
\]

\section{Preliminary observations on open maps}
\label{sec:classification-open}

In this section,
we come back to the notion of "openness" introduced in \Cref{def:open-lambda-term}
and give various examples and counterexamples of open $\lambda$-terms.
We stress the fact that we see the existence of non-open $\lambda$-terms as a great strength of our approach,
as such $\lambda$-terms can still pull back regular languages and allow a greater compositional viewpoint on the logical structures underlying regular languages.

\medskip\noindent
\textbf{Examples of open $\lambda$-terms.}
Statman's finite completeness theorem \cite[Theorem~2]{DBLP:journals/jsyml/Statman82} \cite{https://doi.org/10.48550/arxiv.2309.03602},
a result of the finite model theory of the $\lambda$-calculus,
can be given a geometrical interpretation in our approach.
Indeed, it means that the singleton languages of $\lambda$-terms are regular,
which is equivalent to the fact that $\lambda$-terms are isolated points of the space of profinite $\lambda$-terms,
and also equivalent to the fact that any $\lambda$-term
\[
    M
    \quad\in\quad
    \Lam(1, A)
\]
has "pushforwards".
In other words, all global elements are "open" $\lambda$-terms.

The fact that that the fibration $\Reg \to \Lam$ has simple products implies that the projections
\[
    \pi
    \quad\in\quad
    \Lam(A \times B, A)
\]
are "open" $\lambda$-terms.
In particular, when $A$ is the terminal type $1$,
we get the terminal morphism $!_A : A \to 1$.
The fact that the projections are "open" means that
we have access to existential and universal quantifiers,
computed by
\[
    \push{\pi}
    \ ,\
    \copush{\pi}
    \quad:\quad
    \Reg(A \times B)
    \ \longto\
    \Reg(A)
\]
which are respectively the left and right adjoint to $\pull{\pi}$.

As explained in the work of \Bojanczyk, Klin and Salamanca \cite{Bojaczyk2023},
the fact that surjective letter-to-letter homomorphisms preserve regular languages by direct images is crucial to MSO.
In particular, a word over the alphabet
\[
    \Sigma \times \{0, 1\}^n
\]
amounts to a word together with $n$ predicate on its position,
and formulas of MSO with $n$ second-order variables define languages of words over that alphabet.
The canonical projection
\[
    \Sigma \times \{0, 1\}^{n + 1}
    \ \longto\
    \Sigma \times \{0, 1\}^n
\]
induces a homomorphism of monoids
\[
    (\Sigma \times \{0, 1\}^{n + 1})^*
    \ \longto\
    (\Sigma \times \{0, 1\}^n)^*
\]
whose direct images compute the existential quantifier of MSO
More generally, every homomorphism of monoids $h : \Sigma^* \to \Gamma^*$ preserve regular languages by direct images,
and can be encoded as a $\lambda$-term
\[
    \underline{h}
    \quad\in\quad
    \Lam(\Words{\Sigma}, \Words{\Gamma})
\]
which is "open".
Therefore, the fibrational setting developed in this article encompasses the usual techniques at the heart of the link between regular languages and MSO.

\medskip\noindent
\textbf{Examples of non-open $\lambda$-terms.}
As explained in the introduction,
the prime exemple of non-open $\lambda$-term is the diagonal
\[
    \lambda x. (x, x)
    \quad\in\quad
    \Lam(A, A \times A)
\]
as, in the case of $A = \Nat$, this would contradict the pumping lemma.
Following Lawvere \cite{LawvereEqHyp},
it is customary to define the equality predicate as the "pushforward" of the terminal object of the domain fiber along such a diagonal morphism.
The general unavailability of such a construction in the fibration of regular languages provides evidence for the need to develop new categorical tools better suited to capture the rich logical structure underlying automata theory.

As a consequence of the fact that all global elements are "open" and that some $\lambda$-terms are not "open",
the evaluation $\lambda$-term
\[
    \lambda (f, x). f(x)
    \quad\in\quad
    \Lam((A \To B) \times A, B)
\]
is not "open" itself,
as otherwise all $\lambda$-terms would have "pushforwards",
see \cite[Theorem 3.1]{MelliesZeilberger18}.

\section{Conclusion and future work}
\label{sec:conclusion}
In this paper, we have shown that the class of higher-order regular languages
is closed under the two fundamental constructors of the simply-typed $\lambda$-calculus:
the product type and the arrow type,
as well as under inverse image along $\lambda$-terms.
%
As we explained in the paper,
this amounts to the construction of a "cartesian closed fibration"~$p:\Reg \to \Lam$,
which we carried out in two complementary ways, one fibrational and one topological.
Using that "cartesian closed fibration", we have shown that the notion of Brzozowski derivative generalizes to the higher-order setting.

For future work,
we want to obtain a complete classification of the "open" $\lambda$-terms.
We conjecture that this notion of "openness" may be related to linearity,
as hinted by the fact that the diagonal $\lambda$-term is not "open".
Such a classification would pave the way for a higher-order specification logic,
in the style of MSO for logic on words,
describing exactly the regular languages of $\lambda$-terms.

\bibliography{biblio}

\end{document}